\documentclass[aps,preprint,showpacs,preprintnumbers,amsmath,amssymb,superscriptaddress]{revtex4}

\usepackage{graphicx}
\usepackage{dcolumn}
\usepackage{bm}
\begin{document}


The following article has been submitted to Applied Physics Letters. If it is published, it will be found online at http://apl.aip.org.

\title{Low loss, low dispersion and highly birefringent terahertz porous fibers }

\author{Shaghik Atakaramians}
\affiliation{Centre of Expertise in Photonics and School of Chemistry \& Physics,\\}
\affiliation{Centre for Biomedical Engineering and School of Electrical \& Electronic Engineering,\\ The University of Adelaide, SA 5005, Australia\\}
\author{Shahraam Afshar V.}
\affiliation{Centre of Expertise in Photonics and School of Chemistry \& Physics,\\}
\author{Bernd~M.~Fischer}
\affiliation{Centre for Biomedical Engineering and School of Electrical \& Electronic Engineering,\\ The University of Adelaide, SA 5005, Australia\\}
\author{ Derek Abbott}
\affiliation{Centre for Biomedical Engineering and School of Electrical \& Electronic Engineering,\\ The University of Adelaide, SA 5005, Australia\\}
\author{Tanya M. Monro}
\affiliation{Centre of Expertise in Photonics and School of Chemistry \& Physics,\\}



\begin{abstract}
We demonstrate that porous fibers have low effective material loss over an extended frequency range, 4.5 times larger bandwidth than that can be achieved in sub-wavelength solid core fibers. We also show that these new fibers can be designed to have near zero dispersion for 0.5-1~THz resulting to overall less terahertz signal degradation. In addition, it is demonstrated that the use of asymmetrical sub-wavelength air-holes within the core leads to high birefringence $\approx$~0.026. This opens up the potential for realization of novel polarization preserving fibers in the terahertz regime. \end{abstract}

\maketitle

Bulk optics are used to transport terahertz (THz) radiation in almost all terahertz spectroscopy and imaging systems. However, this limits the integration of THz techniques with infrared and optical system. Although THz waveguides promise to overcome these hurdles, results do date have been limited by the high loss and dispersion. A number of waveguide solutions based on technologies from both electronics and photonics, as reviewed in detail in Ref \cite{Hassani_2008_OEX, Atakaramians_2008_OEX}, have been studied among which solid-core sub-wavelength fibers \cite{Chen_2006_OLT} (called THz microwires~\cite{Afshar_2007_CLEO}), air-core microstructure fibers \cite{Lu_2008_APL} and Ag/PS-coated hollow glass fibers \cite{Bowden_2007_OLT} have the lowest loss reported in the literature for dielectric based waveguides. These fibers are only suitable for relatively narrow band applications.
 
Recently, a novel class of porous fibers for the THz range was suggested independently by two research groups \cite{Hassani_2008_APL,Hassani_2008_OEX,Atakaramians_2008_OEX}. Porous fibers are air-clad fibers with a pattern of sub-wavelength air-holes in the core. Such fibers allow low loss THz propagation and a better confinement of the field relative to microwires.

In this paper, we investigate the frequency dependence of effective material loss and dispersion of THz porous fibers with different sub-wavelength air-hole shapes and compare them to equivalent THz microwires. In the low loss operating regime, we demonstrate a significant improvement in the bandwidth and group velocity dispersion of porous fibers in comparison to THz microwires. Moreover we demonstrate that introducing asymmetrical discontinuities leads to a strong birefringence in the porous fibers avoiding distortion due to polarization mode dispersion. This expands the application of porous fibers to polarization preserving systems, e.g. coherent heterodyne time-domain spectrometry~\cite{Karpowicz_2008_APL}.

\begin{figure}
\centering\includegraphics[height=6cm]{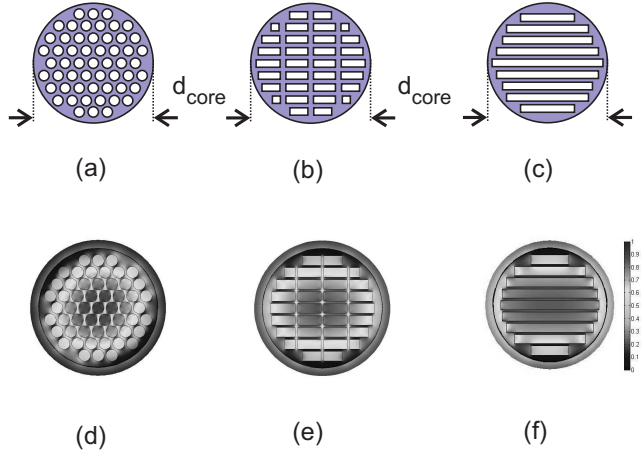}
\caption{\label{fig:cross}Cross-section of porous fibers with (a) circular, (b) rectangular, and (c) slot shaped air-holes. The 2D view of normalized $S_{z}$ for (d) circular shaped air-holes and 57\% porosity, (e) rectangular and 57\% porosity, and (f) slot shaped air-holes and 61\% porosity, at 0.4~THz.}
\end{figure}

The porous fibers studied herein are air-clad and have a pattern of sub-wavelength air-holes within the core. The distribution, shape, and size of the air holes determine the porosity of the structure, which is defined as the fraction of the air-hole area to core area. Field enhancement and localization, occur within these sub-wavelength air-holes as demonstrated in Ref.~\cite{Atakaramians_2008_OEX} and the references therein. In THz porous fibers, the discontinuities are chosen to be air because firstly air is transparent at THz frequencies, has negligible loss, and secondly it gives the highest refractive index contrast, resulting in an increased enhancement of the field. Previously only circular air-holes have been considered, as shown in Fig.~\ref{fig:cross}(a). However, other shapes can also be introduced into the core of these fibers. The degree of enhancement at the air-hole interface depends on the normal component of the electric field, thus in order to have well separated propagation constants for the two polarizations of the fundamental mode, we choose rectangular and slot-shaped sub-wavelength air-holes with the sides of rectangles aligned with the two polarizations. The best arrangement for achieving high porosity values for circular, rectangular, and slot-shaped air-holes are demonstrated in the Fig.~\ref{fig:cross}(a), \ref{fig:cross}(b) and \ref{fig:cross}(c), respectively. Figure~\ref{fig:cross} shows all the three cross-sections and the two-dimensional view of normalized $z$-component of the Poynting vector, $S_{z}$, at 0.4~THz.

\begin{figure}
\includegraphics[height=6cm]{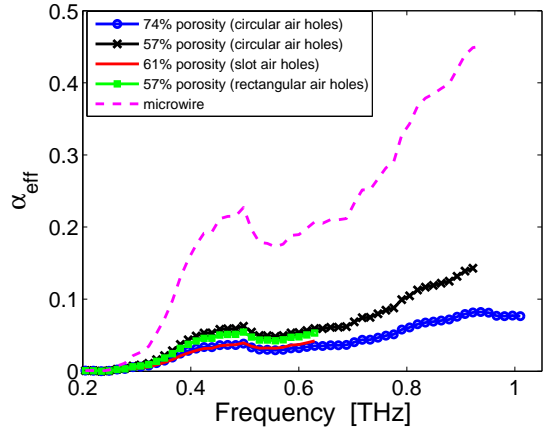}
\caption{\label{fig:Loss} Effective material loss versus frequency of porous fibers with circular shaped air-holes and 74\% porosity and 57\% porosity, porous fibers with rectangular and slot shaped air-holes and 57\% porosity and 61\% porosity, respectively; and a microwire.}
\end{figure}

The effective material loss, $\alpha_{\mathrm{eff}}$, of any mode depends on how the mode is distributed within the inhomogeneous cross section. The calculation of $\alpha_{\mathrm{eff}}$ for porous fibers has been explicitly described in Ref.~\cite{Atakaramians_2008_OEX}. Figure~\ref{fig:Loss} shows the calculated effective material loss of four porous fibers and one microwire. Two porous fibers with circular shaped air-hole and different porosities ($57\%$ and $74\%$), a porous fiber with rectangular shaped air-hole and $57\%$ porosity and a porous fiber with slot shaped air-hole and $61\%$ porosity are chosen for this paper.  The dimension of the fibers are chosen in a way that they have the same amount of loss at 0.2~THz, 0.007~cm$^{-1}$. Cyclic olefin copolymer (TOPAS) is considered as the host material for all of the simulations here. We use the THz properties of TOPAS (refractive index and absorption coefficient as a function of frequency) measured by THz time-domain spectroscopy (THz-TDS) \cite{Fisher_2005_Thesis}. As shown in Fig.~\ref{fig:Loss}, for all the fibers, the loss increases with increasing frequency due to the fact that the power fraction within the core increases. The increase in $\alpha_{\mathrm{eff}}$ of the porous fibers is comparable and less pronounced than in microwire, since less material exists in the core of the porous fibers. Unsurprisingly, the porous fiber with highest porosity has the lowest loss values. For frequencies around 1~Thz, the effective material loss of porous fiber with cicular shaped air-holes and $74\%$ porosity is five times lower than that of the microwire. In this range, 0.2-1~THz, the maximum of the material loss of the porous fiber is 0.08~cm$^{-1}$ while the microwire crosses this value at 0.38~THz. If we take 0.08~cm$^{-1}$ to be a bench mark for the upper limit of the effective material loss, porous fibers offer a factor of 4.5 improvement in the achievable bandwidth.

It should be noted that there is an upper limit, where the effective material loss lines stop in Fig.~\ref{fig:Loss}, beyond which  the fibers either become multi-mode, or for the case of the porous fibers, the material begins to act as an array of independent sub-wavelength fibers. The frequency at which this occurs depends on the porosity of the porous fibers, i.e., for higher porosity values the limit is larger, indicating that the structure stays in the porous fiber single mode regime for a wider range of frequencies. 

\begin{figure}
\includegraphics[height=6cm]{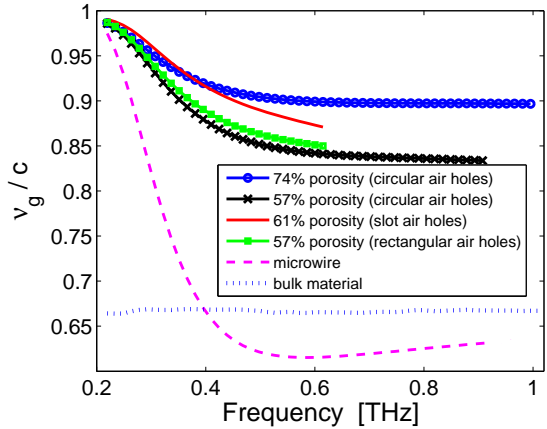}
\caption{\label{fig:VG} Waveguide dispersion versus frequency of porous fibers with circular shaped air-holes and 74\% porosity and 57\% porosity, porous fibers with rectangular and slot shaped air-holes and 57\% porosity and 61\% porosity, respectively; and a microwire.}
\end{figure}

Dispersion is the other mechanism for signal degradation in broadband applications. This occurs when the propagation constant of the guided modes varies with frequency. The frequency dependency of the propagation constant arises from refractive index variation of the host material (material dispersion) or/and waveguide structure (waveguide dispersion) with frequency. The group velocity of the host material, TOPAS, is calculated from the measured refractive index of the bulk material. As shown in Fig.~\ref{fig:VG}, the almost flat feature indicates that the host material has negligible fiber material dispersion. However, for the fibers used here, the waveguide dispersion, which depends on the structure of the fiber, plays an important role. In order to compare the dispersion of the fibers, the group velocity, $\nu_{\mathrm{g}} = \partial\omega/\partial\beta_{\mathrm{eff}}$, is calculated and compared for porous fibers and microwire.

Figure~\ref{fig:VG} shows the group velocity normalized to the speed of light in free space as a function of frequency. For lower frequencies, where the dimension of the fiber is less that the operating wavelength and almost all the power is in the air, the group velocity of the propagating mode in all the structures approaches the velocity of light in free space. By increasing the frequency, the group velocity drops to that of the bulk TOPAS for microwire with a turning point around 0.6~THz (corresponding to zero dispersion), while it drops very slowly to 0.9 and 0.83 for porous fibers with circular shaped air-holes and 71\% and 56\% porosity, respectively, and 0.87 and 0.85 for porous fibers with slot and rectangular shaped air-holes and 61\% and 56\% porosity, respectively, without a turning point. The dispersion of porous fibers with circular shaped air-holes plateaus after 0.5~THz indicating that all the frequency components propagate with constant $\nu_{\mathrm{g}}$, i.e. that dispersion is negligible. Thus terahertz pulse propagating along porous fibers encounters small normal dispersion at the lower frequencies, corresponding to positive chirp in the time domain, and zero dispersion at the higher frequencies, while it encounters strong normal and anomalous dispersion corresponding to positive and negative chirp in the time domain, respectively, in microwires, Fig.~\ref{fig:VG}.

Additional degradation in transmission is associated with birefringence, which can arise from structural and environmental perturbations. This occurs because of different group delays between polarization states, which leads to pulse broadening through polarization mode dispersion. It is well known from optics that the solution to this problem is the use of polarization maintaining (PM) fibers, which introduce modal birefringence into the fiber \cite{Noda_1986_JLT}. Modal birefringence arises from effective refractive index differences between $x$ and $y$ polarization modes, $\left|n_{x}-n_{y}\right|$. Modal birefringence can be introduced using either stress-applying parts in the cladding \cite{Noda_1986_JLT} or/and asymmetry in the core/cladding geometry of the fibers \cite{Suzuki_2001_OEX}. Thus light launched onto one of the principle axes of a PM fiber remains in this polarization in the presence of any environmental perturbations.

\begin{figure}
\includegraphics[height=6cm]{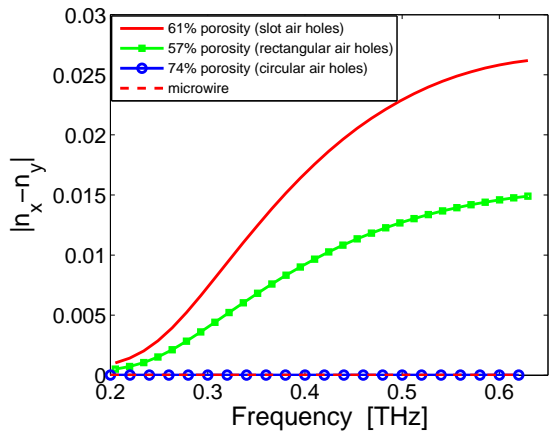}
\caption{\label{fig:DN} Modal birefringence versus frequency of porous fiber with circular shaped air-holes and 74\% porosity , porous fibers with rectangular and slot shaped air-holes and 57\% porosity and 61\% porosity, respectively; and a microwire.}
\end{figure}

Figure~\ref{fig:DN} shows the modal birefringence, $\left|n_{x}-n_{y}\right|$ as a function of frequency. As expected, for symmetrical fibers the birefringence is zero. However, for porous fibers with circular shaped air-holes the calculated birefringence is at the order of $10^{-5}$, nearly zero as shown in Fig.~\ref{fig:DN}. This nonphysical residual birefringence provides a guide to accuracy of the calculation \cite{Atakaramians_2008_OEX}. In contrast for porous fibers with slot and rectangular shaped air-holes, asymmetrical discontinuities in the $x$ and $y$ direction, there is a noticeable birefringence for the fundamental mode. The value of birefringence depends on the shape and arrangement of the holes; the porous fibers with slot and rectangular shaped air-holes proposed in this paper provides a birefringence of 0.026 and 0.015, respectively at 0.6~THz. These values are comparable to recently achieved high birefringence ($\approx 0.025$) in photonic crystals fibers \cite{Chen_2004_OEX}. These high birefringence porous structures have lower single mode operating bandwidth relative to porous fibers with circular shaped air-holes because there is a considerable amount of material between the air-holes and the edge of the core. Recently it has been shown that it is possible to fabricate non-circular air-holes in microstructure optical fibers made up of polymer and soft glasses through extrusion technique \cite{Ebendorff_2007_OEX, Ebendorff_2008_con}, indicating that the fabrication of the proposed structures are feasible.

To conclude, we have presented the effective material loss and group velocity of porous fibers with different sub-wavelength discontinuities as a function of frequency. We showed that there is a factor of 4.5 improvement in the bandwidth over which low loss and low dispersion occurs within the porous fibers compared to microwires. The flexibility of having asymmetrical sub-wavelength discontinuities in the core of porous fibers can lead to a birefringence $\approx$~0.026. Maintaining the polarization of the propagating field in THz waveguides makes these fibers a good substitutes for the free space THz propagation where the polarization state of the THz field is always preserved.

\begin{acknowledgments}
This research was supported under the Australian Research Council's (ARC) \textit{Discovery Projects} funding scheme (project numbers DP0556112 and DP0880436).
\end{acknowledgments}


\end{document}